\PassOptionsToPackage{unicode}{hyperref}
\PassOptionsToPackage{hyphens}{url}
\PassOptionsToPackage{dvipsnames,svgnames,x11names}{xcolor}
\documentclass[
  12pt]{article}

\usepackage{amsmath,amssymb}
\usepackage{iftex}
\ifPDFTeX
  \usepackage[T1]{fontenc}
  \usepackage[utf8]{inputenc}
  \usepackage{textcomp} %
\else %
  \usepackage{unicode-math}
  \defaultfontfeatures{Scale=MatchLowercase}
  \defaultfontfeatures[\rmfamily]{Ligatures=TeX,Scale=1}
\fi
\usepackage{lmodern}
\ifPDFTeX\else  
\fi
\IfFileExists{upquote.sty}{\usepackage{upquote}}{}
\IfFileExists{microtype.sty}{%
  \usepackage[]{microtype}
  \UseMicrotypeSet[protrusion]{basicmath} %
}{}
\makeatletter
\@ifundefined{KOMAClassName}{%
  \IfFileExists{parskip.sty}{%
    \usepackage{parskip}
  }{%
    \setlength{\parindent}{0pt}
    \setlength{\parskip}{6pt plus 2pt minus 1pt}}
}{%
  \KOMAoptions{parskip=half}}
\makeatother
\usepackage{xcolor}
\makeatletter
\ifx\paragraph\undefined\else
  \let\oldparagraph\paragraph
  \renewcommand{\paragraph}{
    \@ifstar
      \xxxParagraphStar
      \xxxParagraphNoStar
  }
  \newcommand{\xxxParagraphStar}[1]{\oldparagraph*{#1}\mbox{}}
  \newcommand{\xxxParagraphNoStar}[1]{\oldparagraph{#1}\mbox{}}
\fi
\ifx\subparagraph\undefined\else
  \let\oldsubparagraph\subparagraph
  \renewcommand{\subparagraph}{
    \@ifstar
      \xxxSubParagraphStar
      \xxxSubParagraphNoStar
  }
  \newcommand{\xxxSubParagraphStar}[1]{\oldsubparagraph*{#1}\mbox{}}
  \newcommand{\xxxSubParagraphNoStar}[1]{\oldsubparagraph{#1}\mbox{}}
\fi
\makeatother

\usepackage{longtable,booktabs,array}
\usepackage{calc} %
\usepackage{etoolbox}
\makeatletter
\patchcmd\longtable{\par}{\if@noskipsec\mbox{}\fi\par}{}{}
\makeatother
\IfFileExists{footnotehyper.sty}{\usepackage{footnotehyper}}{\usepackage{footnote}}
\makesavenoteenv{longtable}
\usepackage{graphicx}
\makeatletter
\def\maxwidth{\ifdim\Gin@nat@width>\linewidth\linewidth\else\Gin@nat@width\fi}
\def\maxheight{\ifdim\Gin@nat@height>\textheight\textheight\else\Gin@nat@height\fi}
\makeatother
\setkeys{Gin}{width=\maxwidth,height=\maxheight,keepaspectratio}
\makeatletter
\def\fps@figure{htbp}
\makeatother

\makeatletter
\@ifpackageloaded{caption}{}{\usepackage{caption}}
\AtBeginDocument{%
\ifdefined\contentsname
  \renewcommand*\contentsname{Table of contents}
\else
  \newcommand\contentsname{Table of contents}
\fi
\ifdefined\listfigurename
  \renewcommand*\listfigurename{List of Figures}
\else
  \newcommand\listfigurename{List of Figures}
\fi
\ifdefined\listtablename
  \renewcommand*\listtablename{List of Tables}
\else
  \newcommand\listtablename{List of Tables}
\fi
\ifdefined\figurename
  \renewcommand*\figurename{Figure}
\else
  \newcommand\figurename{Figure}
\fi
\ifdefined\tablename
  \renewcommand*\tablename{Table}
\else
  \newcommand\tablename{Table}
\fi
}
\@ifpackageloaded{float}{}{\usepackage{float}}
\floatstyle{ruled}
\@ifundefined{c@chapter}{\newfloat{codelisting}{h}{lop}}{\newfloat{codelisting}{h}{lop}[chapter]}
\floatname{codelisting}{Listing}

\makeatother
\makeatletter
\@ifpackageloaded{caption}{}{\usepackage{caption}}
\@ifpackageloaded{subcaption}{}{\usepackage{subcaption}}
\makeatother

\ifLuaTeX
  \usepackage{selnolig}  %
\fi

\usepackage[authoryear]{natbib}
\setcitestyle{aysep={}}

\usepackage{bookmark}

\IfFileExists{xurl.sty}{\usepackage{xurl}}{} %
\definecolor{darkblue}{rgb}{0,0,.3}
\hypersetup{
  pdftitle={What's Old Is New: Improved Learning from Small-Group Tutorials},
  pdfkeywords={randomized experiment, education, teaching probability, artificial intelligence},
  colorlinks=true,
  linkcolor={darkblue},
  filecolor={Maroon},
  citecolor={darkblue},
  urlcolor={darkblue},
  pdfcreator={LaTeX via pandoc}}

\newcommand{\anon}{1}

\if1\anon
  \hypersetup{pdfauthor={John C. Duchi; Will Hartog; Dennis L. Sun}}
\fi

\usepackage{macros}

\begin{document}

\def\spacingset#1{\renewcommand{\baselinestretch}%
{#1}\small\normalsize} \spacingset{1}

\if1\anon
{
  \title{\bf What's Old Is New: Improved Learning from Small-Group Tutorials}
  \author{John C. Duchi \\
    and \\
    Will Hartog \\
    and \\
    Dennis L. Sun \\
    Department of Statistics, Stanford University}
  \maketitle
} \fi

\if0\anon
{
  \bigskip
  \bigskip
  \bigskip
  \begin{center}
    {\LARGE\bf What's Old Is New: Improved Learning from Small-Group Tutorials}
\end{center}
  \medskip
} \fi

\bigskip
\begin{abstract}
The ubiquity of Large Language Models poses a challenge to
student learning in homework-based college courses, where access to
easy homework answers devalues the intended productive struggle.
To
address this concern, we propose Oxbridge-style weekly tutorials, where
students present homework problems without notes in a small group with
an instructor or TA.
We conducted a randomized experiment in an
introductory probability class to evaluate the effectiveness of
tutorials, finding a significant positive effect of tutorial
enrollment on exam grades.
We discuss the implementation, challenges, and
opportunities in this style of course design.
\end{abstract}

\noindent%
{\it Keywords:} randomized experiment, education, teaching probability, artificial intelligence
\vfill

\newpage
\spacingset{1.8} %

\section{Introduction}
\label{sec:intro}

The rise of artificial intelligence, particularly Large Language Models
(LLMs), has upended the role of homework in STEM courses, including
probability and statistics.
Homework has traditionally served several purposes, among them measuring
learning, eliciting effort, and inducing productive struggle and
persistence outside of class.
Growing evidence suggests that LLMs pose a threat to all of these.
First, it appears that LLMs erode the connection between learning and
homework scores, as students can now achieve higher scores without learning
more~\citep{BastaniBaSuGeKaMa25, StrombergLeWu26}.
Second, LLMs reduce cognitive effort and study time, yielding lower-quality
reasoning, downstream problem solving, and even brain
activity~\citep{StadlerBaSa24,RismanchianUzMaCoKu26,
  KosmynaHaYuSiLiBeBrMa25}.
Finally, preliminary work suggests LLMs reduce persistence and hinder
unassisted performance~\citep{LiuChDuBaDu26}.
The traditional approach of assigning and grading homework thus no longer
achieves its intended goals when powerful LLMs can substitute for individual
effort.

Nevertheless, the educational psychology literature demonstrates that
practice testing and distributed practice present two of the most effective
learning strategies~\citep{dunlosky2013improving}.
Homework provides an efficient way to implement both of these strategies,
since it allows students to test themselves outside of class,
spreading studying over time.
It therefore seems premature to abandon homework in the face of LLMs, but
rather to revise assessment so that it achieves its intended goals.

We explore one solution inspired by a centuries-old practice: ``tutorials'' at Oxford
University and ``supervisions'' at Cambridge University, which we collectively call
``Oxbridge-style tutorials.''
Although the particulars vary by school and discipline, these tutorials 
involve small groups of 1--6 students meeting with a ``tutor'' to discuss course
material.
In STEM fields, students typically complete a
set of problems before each tutorial, and the tutor leads the students in a 
discussion based on their answers.

In Section~\ref{sec:tutorials}, we describe how we adapted Oxbridge-style 
tutorials at a private American research university.
We measure the effectiveness of these tutorials for learning using a
randomized experiment in an introductory probability course, which we
discuss in Sections~\ref{sec:design} and \ref{sec:results}. To our knowledge, 
these are the first rigorous quantitative evaluations of the tutorial system.

\section{Tutorials}
\label{sec:tutorials}

Our goal is to retain homework as a method for distributed practice, while migrating
assessment to a setting where AI assistance cannot stand in for a
student's understanding.
We left the weekly problem set essentially unchanged, except that
the write-up no longer carries any weight in the grade.
Instead, students are assessed on a weekly oral presentation in 
a small-group tutorial. 
Therefore, homework becomes preparation for an assessment rather than the
assessment itself---essentially, an instance of learning by preparing to
teach~\citep{Kobayashi18}---and a student who outsources the preparation
pays the price a few days later in the tutorial, rather than several weeks later 
in the exam.
We describe each component of the design in turn.

\subsection{Homework}

We assigned a problem set of six problems each week, which students were required to 
submit, as in a standard course.
For students in the control (non-tutorial) group, we graded homework as usual;
for students in the tutorial group, we recorded submissions
but did not return written feedback.
We continued to require submission for several reasons. 
First, a due date paces work,
meaning it encourages distributed practice rather than cramming before
exams, mimicking the best features of classical
homework~\citep{dunlosky2013improving}.
Second, submission gives a useful commitment device by providing immediate
consequences for not turning problem sets in~\citep{ryan2005effects}.

\subsection{Tutorial Sessions}

Students were assigned to tutorials of 4 students, each led by a teaching
assistant (TA) or the professor, which met for 50 minutes once per week after
the homework due date.
A tutorial consists of 4 problems, one per student, so that every student
presents exactly once.
Importantly, these problems were \emph{variants} of problems from that
week's homework, which the students had not seen before.
The following variants, from the units on counting and on expectation,
are representative.

\begin{example}[Counting and cards]
  In this problem, we evaluate combinatorics using card
  games.
  We tell students to assume cards in the deck are completely randomized.
  \begin{description}
  \item \emph{Homework problem.}
    In Texas Hold'em, each player is dealt 2 cards, and players share 5
    cards in the center of the table.
    A player has a flush if at least 5 of the 7 total cards share a suit.
    What is the probability you get a flush in one hand of Texas Hold'em?
  \item \emph{Tutorial variant.}
    In Bridge, four players play in two-person partnerships, where each
    player is dealt 13 of the 52 cards in the deck.
    A partnership is long in a suit if it has at least 10 cards in that
    particular suit between its two players.
    What is the probability your partnership is long in a suit
    in one hand of Bridge?
  \end{description}
  Both problems require the same decomposition into suits and the same
  summation for the union of (e.g.) 5, 6, or 7 cards of a suit, but no
  verbatim calculation of the Hold'em solution answers the Bridge question.
\end{example}

\begin{example}[Expectations of functions]
  We asked students to compute expectations $\E[g(X)]$ for different
  functions $g$.
  In each variant, $X$ was a random variable taking values in
  $\{2^k\}_{k = 1}^\infty$, with $\P(X = 2^k) = 2^{-k}$ for
  $k = 1, 2, \ldots$.
  \begin{description}
  \item \emph{Homework problem.} Compute $\E[\log_2 X]$.
  \item \emph{Tutorial variant.} Compute $\E[\sqrt{X}]$.
  \end{description}
  Typical misconceptions swap $\E[g(X)]$ for $g(\E[X])$, leading to
  confusion when $\E[X] = +\infty$; a student who has internalized the
  distinction handles either expectation easily.
\end{example}

A student who worked and understood the original problem should find the
variant straightforward; a student who only memorized the solution may
transfer little.

To encourage students to study all problems, tutors did not reveal which
problem a student would present until it was that student's turn,
and students worked at the board without notes.
Those not presenting were asked to stay off devices and to
follow the presenter's argument; when a presenter was stuck, we encouraged
the others to offer suggestions.
In our experience, this inclusion of other students helps to mitigate the
stress of assessment, making clearer that we view the tutorials as something
closer to a working session than examination.

\subsection{Grading}

Perhaps the biggest difference between our implementation of tutorials and 
those at Oxbridge is that our tutorials were graded.
We did this to provide incentives similar to those
of traditional graded homework.
However, the grade was based on perceived effort, engagement, and
participation, rather than mastery (which would be measured on in-class exams).
The evaluation centered on whether they had clearly attempted the homework,
whether they engaged with the problem at the board and with the instructor's
questions, and whether they attended to and helped their peers.
We hope that this evaluation attaches the grade to working the homework
without outsourcing it, measuring the learning process rather than the product.

Instructors recorded a quantized score in $\{0, 1, 2\}$ for each tutorial,
which replaced the homework component of the course grade.
A coarse three-point scale (i) requires only a few seconds to assign, (ii) admits
little argument over boundary cases, and (iii) signals that the
tutorial is not an exam.
A student who prepares but gets stuck on the variant problem receives the
same score as a student who sails through it, encouraging productive struggle.

\subsection{Instructions to Tutors}

We provided tutorial leaders with instructions and materials to standardize
student experience within the tutorial sessions.
The teaching team provided a roadmap for each potential problem in the
weekly problem set consisting of details to ask the students, including
follow-up questions in case time runs long.
Tutorial leaders filled any remaining time with discussion of the other
problems if students finished early.
We also gave tutorial leaders the following general principles:
\begin{itemize}
\item Students should be attentive to their peers when not presenting; they
  should not be reviewing notes or trying to memorize homework.
  As such, the tutorial leader should not let the students know their
  assigned problem until it is their turn to present.
\item The tutorial is not meant to be a quiz.
  It is for building concepts and reinforcing learning, and tutorial grades
  are based not on correctness but preparedness and effort.
\end{itemize}

\section{Experimental Design}
\label{sec:design}

To evaluate the tutorials of Section~\ref{sec:tutorials}, we implemented a
randomized controlled experiment, with blocking based on pretest performance.
We describe the setting, the treatment and control conditions, and our
stratification below.

\subsection{Setting}
Each subject in this study was a student enrolled in Stats \if1\anon {117} \else {XXX}\fi,
Introduction to Probability, in Spring 2026.
The course covers concepts in a standard probability course, with
major units on foundations of probability, counting, discrete random
variables, expectation and variance, conditioning, and continuous 
random variables.
The 10-week course consisted of three 50-minute lectures per week, 
with one problem set assigned as homework each week.
Students in both conditions attended the same lectures and sat the same
in-class midterm and final examinations, which supply the outcome measures
for the analysis of Section~\ref{sec:results}.

\subsection{Treatment group}
Students assigned to the treatment group participated in the weekly
tutorials of Section~\ref{sec:tutorials}.
As described there, they still handed in their written homework, but it was
not graded; they were evaluated exclusively on their tutorial performance,
which took the place of the homework component of their course grade.

\subsection{Control group}
Students in the control group participated in an essentially standard
college-level course, at least as offered in the United States.
They turned in written homework for grading, and the
course staff advised them against using AI assistance, suggesting that
it was likely to hinder their performance on the midterm and final exams.
Weekly 50-minute sections were offered by TAs, and students were 
asked to attend one section each week.
These sections consisted of live-solving of problems involving either
concepts or examples from the weeks' lectures, led by TAs.
Students who attended section each week spent exactly the same time
as students in tutorials.
However, we did not enforce attendance in these sections, so students in the control 
group who chose not to attend had less time-on-task than students in the 
treatment group.

\subsection{Blocking and control}

To reduce variation due to differences in student background, 
we blocked based on performance
on a concept inventory pre-test, which was issued
on the first day of class.
This inventory consisted of 12 total questions that evaluated
students' conceptual and intuitive understanding of probability, 
related to (i) independence, (ii) sequential sampling (geometric random variables), 
(iii) sampling with and without replacement, (iv) expected values, (v) and conditioning.

We then blocked students based on their pretest score and randomized students 
with the same score so that approximately half were in treatment and half were in 
control.
So, for example, among the $5$ students scoring 12/12 on the pretest, 3 were
in tutorials and 2 were in the control group, and among the 10 students
scoring 9/12 on the pretest, 5 were in the tutorials and 5 in the control.

\section{Results and analysis}
\label{sec:results}

We present summary results in this section for students who consented 
to have their data included in the study.
Of the $80$ students who enrolled in the class, all 
$n=66$ who responded to the form consented to the use of their data.
The remaining students who did not respond may result in some
selection bias; while we cannot present the full results, they remain
consistent with the conclusions here.
We present two analyses: a standard fixed-effects analysis of a
randomized block design and a fully nonparametric randomization test.
All inferences ($p$-values, confidence intervals) are one-sided.

\subsection{Fixed-effects analysis of a randomized block design}

Let $W_i \in \{ 0, 1 \}$, $X_i \in \{-1, 0, 1, \dots, 12 \}$, and $Y_i$
denote the treatment assignment, pretest score, and outcome (e.g., midterm
or final score), respectively, of student $i$, where
we code $X_i = -1$ to indicate that the student did not
complete the pretest.
We fit the standard linear model
\begin{equation}
   Y_i = \tau \indic{W_i = 1} + \sum_{k=-1}^{12} \beta_k \indic{X_i = k}
   \label{eq:lm_block}
\end{equation}
to the exam scores.
In Model~\eqref{eq:lm_block}, the average treatment effect $\tau$
constitutes the parameter of interest,
and our randomization guarantees that least squares
is consistent for this quantity~\citep{ImbensRu15}.
Table~\ref{table:factor} summarizes the results, showing estimated
treatment effects of 14.9 percentage points for the midterm and 
11.4 percentage points for the final
($p < 0.002$), with 95\% lower confidence bounds for the effect size
$\tau$ (in percentage points) of 6.84 percentage points the midterm
and 5.30 percentage points on the final.

\begin{table}
  \begin{center}
    \begin{tabular}{|c|c|c|c|c|c|c|}
      \hline
      Outcome & $n$ & $\what{\tau}$ (pp) & std.\ error (pp) & t-score & $p$-value & 95\% lower
      $\tau$ (pp) \\ \hline\hline
      Midterm & 66 & 14.9 & 4.84 & 3.08 & 0.002 & 6.84 \\
      \hline
      Final & 64 & 11.4 & 3.66 & 3.12 & 0.001 & 5.30 \\
      \hline
    \end{tabular}
  \end{center}
  \caption{\label{table:factor} Results of the fixed-effects analysis
    by fitting Model~\eqref{eq:lm_block}, measuring effect sizes in
    percentage points.
    Two students dropped the course after the midterm examination and did
    not take the final.}
\end{table}

\subsection{Randomization test}

Under the null hypothesis that the treatment has no effect, the distribution
of any test statistic is invariant to permutations of the treatment
assignments.
Thus, the permutation null---permuting treatment assignments and
recalculating the test statistic---provides a reference null distribution
for any test.
(We permute treatment assignments separately for each pretest score to
mirror the randomized block design of the study.)
One obtains a $p$-value by locating the observed difference within this
reference null distribution.

So long as the test statistic is monotone in outcome differences, the
randomization test remains valid for testing the one-sided null hypothesis
\begin{equation}
  \label{eqn:fisher-null}
  H_0 : Y_i(1) \le Y_i(0) ~~ \mbox{for~each~} i,
\end{equation}
where $Y_i(t)$ denotes the {\em potential outcome}
of subject $i$ under treatment $t$~\citep{ImbensRu15}.
Under the null hypothesis~\eqref{eqn:fisher-null}, participation in
tutorials has no positive effect on any individual's midterm or final
performance, and this $p$-value is exact, without any assumptions on the
distribution of the potential outcomes.
Following \citet{CaugheyDaLiMi23}, we also identify the smallest $\tau$ for
which the hypothesis
\begin{align}
  \label{eqn:fisher-null-with-c}
  H_{\tau} : Y_i(1) \le Y_i(0) + \tau ~~ \mbox{for each}~ i
\end{align}
cannot be rejected at level $\alpha = .05$.
This provides the natural analog of the ``95\% lower confidence bound'' in
the fixed-effects analysis.

Table~\ref{table:randomization-results} summarizes results, yielding
estimated treatment effect (difference in means between the two
groups) of 14.3 percentage points for the midterm and 9.5 percentage
points for the final ($p < 0.002$).
The ``95\% lower confidence bound'', in the sense
of~\eqref{eqn:fisher-null-with-c}, is 6.78 percentage
points and 5.19 percentage points.

\begin{table}[ht]
  \centering
  \begin{tabular}{|c|c|c|c|c|c|c|}
    \hline
    Outcome & $n$ & Tutorial mean & Control mean & $\what{\tau}$ (pp) & $p$-value & 95\% lower $\tau$ (pp) \\
    \hline\hline
    Midterm & 66 & 72.7\% & 58.4\% & 14.3 & 0.002 & 6.78 \\
    \hline
    Final & 64 & 59.5\% & 50.0\% & 9.5 & 0.002 & 5.19 \\
    \hline
  \end{tabular}    
  \caption{\label{table:randomization-results} Results of the randomization
    test in percentage points.}
\end{table}

\section{Conclusions}
\label{sec:conclusions}

The significant positive effects we observe in our randomized 
experiment suggest that tutorials are a potentially powerful educational 
invention.
However, we cannot identify the precise causal mechanism in this study.
We hypothesize that at least two factors coincide here: in addition to
the personalized instruction that tutorials afford, AI may also be depressing 
learning from traditional homework.
If this hypothesis is correct, then tutorials are not just a useful
educational intervention but one that is uniquely suited to mitigating the
detrimental effects of AI on learning.

A natural concern that tutorials raise is their cost: how is it possible to 
achieve ratios of 4 students per TA? In order for tutorials to be feasible, 
a university needs to have ratios of under 4 students per TA-hour. 
The ratio needs to be under 4 to allow time for preparation and administrative
duties, but tutorials also reduce the
need for other traditional TA duties, such as grading and 
office hours (since students have the opportunity to ask questions in
tutorial).
At our private research university, the ratio is 1.5--2.0,
but this ratio varies widely across institutions.
At large public
research universities, a brief survey
suggests the ratio ranges from roughly 3.0, at the
University of Michigan~\citep{Gala-Garza24}, UT
Austin~\citep[Tab.~2]{McMillan26}, and UCLA~\citep[``Secondary
  Sections'']{OLearySaShGaCoLeEaVa21},
to 4.0 at UCSD~\citep[Sec.~4.5]{UCSDStaffing26} and \citet{OregonStaffing26}.

As a practical matter, even if a TA with a 20-hour-per-week appointment 
could technically hold 15 tutorials, that may not be desirable. Our TAs had
shorter, 10-hour-per-week appointments, meaning that each TA only needed
to hold 4 tutorials each week (i.e., 2 days with 2 tutorials each). We have 
also had success hiring senior undergraduates to serve as tutorial leaders, 
who excelled in a previous offering of the class and were 
recommended by their tutorial leader. All of these strategies can make 
tutorials more feasible when TA funding is limited.

Many questions remain.
Most saliently, our study focuses on a single course on introductory
probability at one university, with a single instructor.
Because we encouraged, but did not enforce, the control group to attend
sections, time on task or additional time with instructors may influence the
observed improvements.
Moreover, our study does not disentangle more precise causative mechanisms:
the tutorials include unseen variant problems, require explaining solutions
at the board, allow small-group participation, and use effort-based
grading.
Larger classes may necessitate changes to the format, such as asynchronously
scheduled one-on-one or remote sessions, and it is unclear whether the
observed benefits would survive these changes.
We are currently investigating these questions in a larger study across
several courses in several different disciplines.

Even with these caveats, our results suggest that tutorials provide
a promising response to new technology.
For centuries, tutors at Oxbridge have asked students to explain their work,
and when a written solution may no longer reflect what a student
understands, this approach may provide precisely those incentives that
keep students learning.
While an LLM can type a student's homework, it cannot put chalk to the
chalkboard for them.

\section{Disclosure statement}\label{disclosure-statement}

The authors have no conflicts of interest.

The authors used Claude Opus 5.5 to perform the analyses in this paper and 
to edit the writing.

\section{Data Availability Statement}
\label{data-availability-statement}

Because of the sensitivity of personally identifiable information in the
analyses, we did not receive permission from the Institutional Review Board 
to make the original data available.
Nevertheless, we make our full analysis pipeline available below.
\if1\anon
This study was approved by the Stanford Institutional Review Board,
IRB \#83954.
\fi

\appendix

\section{Analysis details}
\label{sec:analysis-details}

\newcommand{\Tobs}{T_{\textup{obs}}}
\newcommand{\Tpi}[1][]{
  \ifthenelse{\isempty{#1}}{%
    T_{\pi}
  }{%
    T_{\pi({#1})}
  }
}
\newcommand{\bZ}{\boldsymbol{Z}}
\newcommand{\obs}{^{\textup{obs}}}

\subsection{Analysis of the randomization tests}

For observed scores $Y_i\obs$ and a vector $\bZ = (Z_i)_{i = 1}^n \in \{0,
1\}^n$ of treatment indicators and associated counts $n_z = \sum_{i=1}^n
\indic{Z_i = z}$, $z = 0, 1$, we consider the sample mean difference
\begin{align*}
  T_{\bZ} \defeq \frac{1}{n_1} \sum_{i : Z_i = 1} Y_i\obs
  - \frac{1}{n_0} \sum_{i : Z_i = 0} Y_i\obs,
\end{align*}
so that for the actual assigments $\bZ\obs$ we have $T_{\bZ\obs} =
\frac{1}{n_1} \sum_{i : Z_i\obs = 1} Y_i(1) - \frac{1}{n_0} \sum_{i :
  Z_i\obs = 0} Y_i(0)$.
To evaluate the null~\eqref{eqn:fisher-null}, we consider all randomizations
$\bZ^* \in \{0, 1\}$ respecting the pretest scores and initial counts, so
that $\bZ^*$ are sampled uniformly over the collection of treatment
assignments with number of individuals in each pre-test strata equal to that
in $\bZ\obs$.
As \citet{CaugheyDaLiMi23} show,
the randomization probability
$\P(T_{\bZ\obs} \le T_{\bZ^*} \mid Y\obs)$ is a valid
$p$-value for the weaker one-sided null
that $Y_i(1) \le Y_i(0)$ for each $i$.
We also identify the smallest value $\tau$ at
which we can reject the null

We estimate the value $\what{\tau}$ in the null~\eqref{eqn:fisher-null-with-c}
at level $\alpha$ as follows.
Imputing at the boundary $Y_i(1) = Y_i(0) + \tau$ yields $Y_i(0)
= Y_i\obs + \tau Z_i\obs$ and $Y_i(1) = Y_i\obs - \tau (1 - Z_i\obs)$, giving
the adjusted statistic
\begin{align*}
  T_{\bZ, \tau} \defeq T_{\bZ} + \tau \left(\frac{1}{n_1}
  \<\bZ\obs, \bZ\> - \frac{1}{n_0} \<\bZ\obs, \ones - \bZ\>\right)
  = T_{\bZ}
  - \tau \left(1
  + \left(\frac{1}{n_1} - \frac{1}{n_0}\right) \<\bZ\obs, \bZ\>\right).
\end{align*}
\citet{CaugheyDaLiMi23} show that
$\P(T_{\bZ\obs} \le T_{\bZ^*,\tau} \mid Y\obs)$ is a valid
$p$-value for the null~\eqref{eqn:fisher-null-with-c}.
To compute $p$-values, we use a Monte-Carlo simulation
of $N = 10^6$ resamples of $\bZ$,
including $\bZ\obs$ itself, 
computing the one-sided $p$-value
$p = \frac{1}{N} \sum_{i = 1}^N \indic{T_{\bZ\obs} \le T_{\bZ_i^*}}$.
Because $\tau \mapsto T_{\bZ, \tau}$ is monotonically decreasing,
we can efficiently solve for $\what{\tau}$, the largest $\tau$ at which
we can reject the null~\eqref{eqn:fisher-null-with-c}
at level $\alpha$ by solving
\begin{align*}
  \what{\tau}
  \defeq \sup\Big\{\tau
  \mid \alpha > \frac{1}{N} \sum_{i = 1}^N \indic{T_{\bZ\obs}
    \le T_{\bZ^*_i, \tau}}
  \Big\}.
\end{align*}

\phantomsection\label{supplementary-material}
\bigskip

\begin{center}

{\large\bf SUPPLEMENTARY MATERIAL}

\end{center}

The full analysis pipeline of our results are available publicly 
on Github at 

\if1\anon
{
\begin{center}
{\tt https://www.github.com/dlsun/tutorials}.
\end{center}
} 
\else
{
\begin{center}
{\tt REDACTED}.
\end{center}
} 
\fi

We did not receive permission to make the original data available, so the 
data files in the repository all contain simulated data that is structured 
exactly like our original data. However, in order to permit
validation of our results by approved individuals, we include
the following SHA256 hashes of our original data files, also
described in the \texttt{PIPELINE.md} file there:
\begin{itemize}
\item \texttt{raw-sp26-gradescope-grades.csv} \\
  ac077d0483a4b4ef36be07fc6073e65224df938c658ae7bf2a3226da09d1e727
\item \texttt{raw-tutorial-assignments.csv} \\
  1c7d7e0cadca32c7c90763e19c0bcdb48324747ac8760828a6ecb0dd0d83f7d5
\item \texttt{raw-consent.csv} \\
  1d89697dcd90b76c3244997991e135c0b31a9856d09f67df450003931f76693c
\item \texttt{raw-pretest-scores.csv} \\
  a051dea5c5aec15b36c204285a2b3a8b71219e74e7cc85c1f4e697eb9c40686b
\end{itemize}

\bibliography{bib,bibliography}

\begin{thebibliography}{16}
\providecommand{\natexlab}[1]{#1}
\providecommand{\url}[1]{\texttt{#1}}
\expandafter\ifx\csname urlstyle\endcsname\relax
  \providecommand{\doi}[1]{doi: #1}\else
  \providecommand{\doi}{doi: \begingroup \urlstyle{rm}\Url}\fi

\bibitem[Bastani et~al.(2025)Bastani, Bastani, Sungu, Ge, \"{O}zge Kabakc\i,
  and Mariman]{BastaniBaSuGeKaMa25}
H.~Bastani, O.~Bastani, A.~Sungu, H.~Ge, \"{O}zge Kabakc\i, and R.~Mariman.
\newblock Generative {AI} without guardrails can harm learning: Evidence from
  high school mathematics.
\newblock \emph{Proceedings of the National Academy of Sciences}, 122\penalty0
  (26):\penalty0 e2422633122, 2025.

\bibitem[Caughey et~al.(2023)Caughey, Dafoe, Li, and Miratrix]{CaugheyDaLiMi23}
D.~Caughey, A.~Dafoe, X.~Li, and L.~Miratrix.
\newblock Randomisation inference beyond the sharp null: bounded null
  hypotheses and quantiles of individual treatment effects.
\newblock \emph{Journal of the Royal Statistical Society, Series B},
  85:\penalty0 1471--1491, 2023.

\bibitem[Dunlosky et~al.(2013)Dunlosky, Rawson, Marsh, Nathan, and
  Willingham]{dunlosky2013improving}
J.~Dunlosky, K.~A. Rawson, E.~J. Marsh, M.~J. Nathan, and D.~T. Willingham.
\newblock Improving students’ learning with effective learning techniques:
  Promising directions from cognitive and educational psychology.
\newblock \emph{Psychological Science in the Public interest}, 14\penalty0
  (1):\penalty0 4--58, 2013.

\bibitem[Gala-Garza(2024)]{Gala-Garza24}
T.~Gala-Garza.
\newblock {LSA} reduces {GSI} allocations by 8\% for fall 2024.
\newblock \emph{The Michigan Daily}, 2024.
\newblock URL
  \url{https://www.michigandaily.com/news/academics/lsa-reduces-gsi-allocations-by-8-for-fall-2024/}.

\bibitem[Imbens and Rubin(2015)]{ImbensRu15}
G.~Imbens and D.~Rubin.
\newblock \emph{Causal Inference for Statistics, Social, and Biomedical
  Sciences}.
\newblock Cambridge University Press, 2015.

\bibitem[Kobayashi(2019)]{Kobayashi18}
K.~Kobayashi.
\newblock Learning by preparing-to-teach and teaching: A meta-analysis.
\newblock \emph{Japanese Psychological Research}, 61\penalty0 (3):\penalty0
  192--203, 2019.

\bibitem[Kosmyna et~al.(2025)Kosmyna, Hauptmann, Yuan, Situ, Liao, Beresnitzky,
  Braunstein, and Maes]{KosmynaHaYuSiLiBeBrMa25}
N.~Kosmyna, E.~Hauptmann, Y.~T. Yuan, J.~Situ, X.-H. Liao, A.~V. Beresnitzky,
  I.~Braunstein, and P.~Maes.
\newblock Your brain on {ChatGPT}: Accumulation of cognitive debt when using an
  {AI} assistant for essay writing task.
\newblock \emph{arXiv:2506.08872 [cs.AI]}, 2025.

\bibitem[Liu et~al.(2026)Liu, Christian, Dumbalska, Bakker, and
  Dubey]{LiuChDuBaDu26}
G.~Liu, B.~Christian, T.~Dumbalska, M.~A. Bakker, and R.~Dubey.
\newblock {AI} assistance reduces persistence and hurts independent
  performance.
\newblock \emph{arXiv:2604.04721 [cs.AI]}, 2026.

\bibitem[McMillan(2026)]{McMillan26}
A.~McMillan.
\newblock Soft money instructional funds allocation guidelines.
\newblock The University of Texas at Austin, College of Liberal Arts, 2026.
\newblock URL
  \url{https://utexas.atlassian.net/wiki/spaces/larc/pages/75241069/Soft+Money+Instructional+Funds+Allocation+Guidelines}.
\newblock Accessed October 14, 2026.

\bibitem[O'Leary et~al.(2021)O'Leary, Sayson, Shapiro, Garfinkel, Conley,
  Levis-Fitzgerald, Eagan, and Van~Valkenburgh]{OLearySaShGaCoLeEaVa21}
E.~S. O'Leary, H.~W. Sayson, C.~Shapiro, A.~Garfinkel, W.~J. Conley,
  M.~Levis-Fitzgerald, M.~K. Eagan, and B.~Van~Valkenburgh.
\newblock Reimagining the introductory math curriculum for life sciences
  students.
\newblock \emph{CBE---Life Sciences Education}, 20\penalty0 (4):\penalty0 ar62,
  2021.
\newblock \doi{10.1187/cbe.20-11-0252}.
\newblock URL \url{https://doi.org/10.1187/cbe.20-11-0252}.

\bibitem[{Oregon State University, College of Earth, Ocean, and Atmospheric
  Sciences}(2026)]{OregonStaffing26}
{Oregon State University, College of Earth, Ocean, and Atmospheric Sciences}.
\newblock {CEOAS GTA} allocation guidelines, 2026.
\newblock URL \url{https://ceoas.oregonstate.edu/gta-allocation-guidelines}.
\newblock Accessed October 14, 2026.

\bibitem[Rismanchian et~al.(2026)Rismanchian, Uzun, Matayoshi, Cosyn, and
  Kurd-Misto]{RismanchianUzMaCoKu26}
S.~Rismanchian, H.~Uzun, J.~Matayoshi, E.~Cosyn, and E.~Kurd-Misto.
\newblock Faster completion, less learning: Generative {AI} reduced study time
  on math problems and the knowledge they build.
\newblock \emph{arXiv:2605.21629 [cs.CY]}, 2026.

\bibitem[Ryan and Hemmes(2005)]{ryan2005effects}
C.~S. Ryan and N.~S. Hemmes.
\newblock Effects of the contingency for homework submission on homework
  submission and quiz performance in a college course.
\newblock \emph{Journal of Applied Behavior Analysis}, 38\penalty0
  (1):\penalty0 79--88, 2005.
\newblock \doi{https://doi.org/10.1901/jaba.2005.123-03}.
\newblock URL
  \url{https://onlinelibrary.wiley.com/doi/abs/10.1901/jaba.2005.123-03}.

\bibitem[Stadler et~al.(2024)Stadler, Bannert, and Sailer]{StadlerBaSa24}
M.~Stadler, M.~Bannert, and M.~Sailer.
\newblock Cognitive ease at a cost: {LLM}s reduce mental effort but compromise
  depth in student scientific inquiry.
\newblock \emph{Computers in Human Behavior}, 160\penalty0 (108386):\penalty0
  1--7, 2024.

\bibitem[Str\"{o}mberg et~al.(2026)Str\"{o}mberg, Lei, and Wu]{StrombergLeWu26}
D.~Str\"{o}mberg, V.~Lei, and Y.~Wu.
\newblock The generative {AI} learning penalty: Evidence from {C}hinese
  secondary education.
\newblock \emph{Centre for Economic Policy Research}, \penalty0 (DP21577),
  2026.
\newblock Discussion Paper.

\bibitem[{Workgroup on the Future of Graduate Education}(2026)]{UCSDStaffing26}
{Workgroup on the Future of Graduate Education}.
\newblock Final report.
\newblock Technical report, University of California, San Diego, June 2026.
\newblock URL
  \url{https://senate.ucsd.edu/media/768814/sawg-on-future-of-graduate-education-review-docs.pdf}.

\end{thebibliography}

\end{document}